\documentclass[aps,twocolumn,prb,showpacs]{revtex4}

\usepackage{epsfig}

\begin{document}

\title{Optical asymptotics via Weniger transformation}

\author{Riccardo Borghi}

\address{Dipartimento di Elettronica Applicata, Universit\`a Roma Tre, 
and CNISM \\
Via della Vasca Navale 84, I-00146 Rome, Italy\\
{\tt e-mail: borghi@uniroma3.it}
}

\begin{abstract}
Starting from the resurgence equation discovered by Berry and Howls [M. V. Berry and C. Howls,
``Hyperasymptotics for integrals with saddles," Proc. R. Soc. Lond. A {\bf 434,} 657-675 (1991)], the
Weniger transformation is here proposed as a natural, efficient, and straightforwardly implementable 
scheme for the efficient asymptotics evaluation of a class of integrals occurring in several areas of 
physics and, in particular, of optics. Preliminary numerical tests, carried out on the Pearcey function, 
provide a direct comparison between the performances of Weniger transformation and those of Hyperasymptotics,
which seems to corroborate the theoretical predictions. We believe that Weniger transformation would be 
a very useful computational tool for the asymptotic treatment of several optical problems.
\end{abstract}


\maketitle

Physical asymptotics finds a natural field of applications in optics. 
It is well known, in fact, that wave optics usually reduces to geometrical optics
when the wavelength of the radiation is let to vanish, and that the
result of such limiting procedure turns out to be singular at special loci, 
called caustics,  where unphysical intensity divergences are predicted.\cite{born&wolf} 
The  above transition leads to a  mathematical description of the wavefield
given in terms of series expansions which turn out to be divergent.\cite{hardy}
Despite the opinion of the great mathematician Abel,\cite{abel} 
divergent series are fashinating objects which  are considered 
tools of invaluable importance in physics.
Within the asymptotics realm, diverging series naturally arise when the 
highly oscillating integrals representing the wavefields are 
 evaluated via standard asympotic methods, like the steepest-descent.\cite{born&wolf}
While it is a well known fact that the leading asymptotic expansion 
is given  by the sum of all the contributions coming from the different saddle points
 belonging to the integration path,\cite{born&wolf}  it is less known  
 that all higher-orders asymptotic corrections can be sistematicly evaluated and arranged in the form of a   
 diverging series.\cite{dingle}  Berry and Howls (BH henceforth)\cite{berryPRSA-90,berryPRSA-91} 
 developed an interpretation scheme of the above divergence, termed \emph{hyperasymptotics} (H 
 henceforth), in which the  divergent character of the asymptotic 
 series obtained from the Dingle's procedure is regarded as a source of precious information,
 to be suitably decoded, rather than a mathematical pathological behavior. 
 The central result of BH is that such a  decoding can be achieved by taking into account an intricate 
 sequence of truncated asymptotic series involving the \emph{whole} network of saddles, including also those
 which do not belong to the integration path.
 Since its introduction, H has been estensively used, up to recent time.\cite{alvarezJPA-02,parisJCAM-06}

 The aim of the present Letter is to show that the decoding operated by H can also be achieved
 by using a nonlinear resummation scheme, the so-called Weniger transformation (WT henceforth),\cite{wenigerCPR-89} 
 when is let acting directly on the diverging series expansions built up only at the saddle points belonging at the steepest descent 
 path. In particular, our idea of employing WT is based on the universal  ``factorial divided by power" law, ruling the asymptotic
 behavior of the Dingle's expanding coefficients, which naturally  arises 
 from the H scheme, and the fact that the same type
 of divergence is displayed by the Euler series, for which WT has been proved to be the most efficient resummation scheme.\cite{wenigerCPR-89}  
 We also present a direct numerical  comparison between the results obtained 
 by H and by WT when the wavefield associated with the so-called cusp diffraction catastrophe,\cite{berryPIO-80} mathematically 
 described by the Pearcey function, is asympotically evaluated. 
 
First of all, we briefly resume the main results of the Dingle's work and 
of its refinements obtained by BH. 
For simplicity we shall limit ourselves to the following class of phase 
integrals:
\begin{equation}
\mathcal{I}=\displaystyle\int_{\mathcal{C}}\,\exp[\mathrm{i}\,\Phi(s)]\,\mathrm{d}s,
\label{generic}
\end{equation}
where $\Phi(s)$ is a function of the variable $s$ and  $\mathcal{C}$ denotes a steepest-descent integration path, in the complex plane of $s$, passing through one or more saddle points, denoted by $\{s_n\}$, which are solutions of the equation $\mathrm{d}\Phi({s})/\mathrm{d}s=0$. The integral in Eq.~(\ref{generic}) is then customarily written as the sum of separate contributions, say $\mathcal{I}_k$, corresponding to the saddle, $s_k$, belonging to the 
integration path, formally expressed as\cite{dingle,berryPRSA-91}
\begin{equation}
\mathcal{I}_k =
\exp(\mathrm{i}\Phi_k)\,\displaystyle\sum_{r=0}^\infty\,T^{(k)}_r,
\label{Dingle.1}
\end{equation}
 where $\Phi_k=\Phi(s_k)$ and
\begin{equation}
T^{(k)}_r=
(r-1/2)!\,\,\,\mathrm{i}^{r+1/2}\,
\displaystyle\frac{1}{2\pi\mathrm{i}}\,
\displaystyle\oint_{\Gamma_k}\,\displaystyle\frac{\mathrm{d}s}{[\Phi(s)-\Phi_k]^{r+1/2}},
\label{Dingle.2}
\end{equation}
with $\Gamma_k$ denoting a small positive loop around the saddle $s_k$.
The fundamental result found by BH was that, although in principle only the saddle $s_k$
does contribute to the integral in Eq.~(\ref{Dingle.1}), 
all other saddles $s_{h}$, with $h \ne k$, thus including those not 
belonging to the path $\mathcal{C}$, are involved through  the following, formally 
exact, \emph{resurgence} formula:\cite{berryPRSA-91}
\begin{equation}
T^{(k)}_r=
\displaystyle\frac 1{2\pi\mathrm{i}}\,
\displaystyle\sum_{h=1 \atop h\ne k}^{\rm all\,\,\,saddles}\,
(-1)^{\gamma_{kh}}\,
\displaystyle\sum_{t=0}^\infty\,
\displaystyle\frac{(r-t-1)!}{F^{r-t}_{kh}}\,T^{(h)}_t,
\label{resurgence}
\end{equation}
where the (complex) quantities $F_{kh}=\Phi_k-\Phi_h$, termed \emph{singulants},
were introduced, and $\gamma_{kh} \in \{0,1\}$ are binary quantities,
related to the topology of the saddles distribution.\cite{berryPRSA-91}
The resurgence effect in Eq.~(\ref{resurgence}) constituted the
basis for the development of H, and the same equation will be the starting point of our analysis. 
 Equation~(\ref{resurgence}) 
gives the  
asymptotic law of $T^{(k)}_{r}$ for large $r$, i.e.,\cite{berry234}
\begin{equation}
T^{(k)}_r \propto
\displaystyle\frac{(r-1)!}{F^r_{kh^*}},
\label{resurgence.2}
\end{equation}
where $h^*$ indicates the saddle corresponding to the minimum value of $|F_{kh^*}|$.
In particular, Eq.~(\ref{resurgence.2}) explains why the adiancent saddles are responsible of the divergent
character of the series in Eq.~(\ref{Dingle.1}). The ``factorial divided by power" asymptotic law in 
Eq.~(\ref{resurgence.2}) is also related to the fact that the coefficients $T^{(k)}_r$ 
are basically generated via a local expansion involving high derivatives at the saddle $s_k$, a feature which is 
common to all asymptotic methods.\cite{dingle,berry234} 
In particular, Eq.~(\ref{resurgence.2}) shows that the divergent 
character of the Dingle' series in Eq.~(\ref{Dingle.2}) is equivalent to that of 
the so-called Euler series, defined as\cite{erdelyi}
\begin{equation}
\displaystyle\sum_{r=0}^\infty\,z^r\,r!=
\displaystyle\int_0^\infty\,\displaystyle\frac{\exp(-t)}{1-zt}\,\mathrm{d}t,
\label{resurgence.2.1.1}
\end{equation}
which, on the other hand, is strictly related to the functional form of 
some quantities, called \emph{hyperterminants}, which are of pivotal 
importance in H.\cite{berryPRSA-91} 
Weniger showed that the Euler series in Eq.~(\ref{resurgence.2.1.1}) can be 
summed directly from the sequence of its partial sums.\cite{wenigerCPR-89}
What is more important, WT has proved to be the most efficient way for 
the resumming to be achived when compared to other resummation schemes, like for instance those 
involving Pad\'e approximants.\cite{wenigerCPR-89,baker} 
We do not wish entering into the mathematical details about WT, for 
which we suggest the interested reader to consult the extensive literature (see for 
instance the recent review in Ref.~\onlinecite{wenigerJMP-04} and 
Ref.~\onlinecite{wenigerArXiv}). Our aim is rather to give 
a first numerical evidence about the retrieving capabilities of WT when is applied
for the asymptotic evaluation of a class of integrals of fundamental importance in optics,
which have been treated with the use of H. This will also deserve to provide  a numerical comparison between H and WT.

Basically, WT operates a transformation on the sequence of the 
partial sums of a diverging series to convert it into a new sequence which 
should converge to the generalized  limit of the series itself.
In our case WT acts on the sequence, $\{S^{(k)}_{m}\}$, of the partial sums of the series in Eq.~(\ref{Dingle.1}), 
and converts it into a new sequence, say $\{\delta^{(k)}_m\}$, defined by\cite{wenigerCPR-89}
\begin{equation}
\delta^{(k)}_m=\displaystyle\frac
{\displaystyle\sum_{j=0}^m\,(-1)^j \left(m\atop j\right)\,(1+j)_{m-1} 
\displaystyle\frac{S^{(k)}_j}{T^{(k)}_{j+1}}}
{\displaystyle
\sum_{j=0}^m\,(-1)^j \left(m\atop j\right)\,(1+j)_{m-1} 
\displaystyle\frac{1}{T^{(k)}_{j+1}}}.
\label{weniger}
\end{equation}
The form of WT given in Eq.~(\ref{weniger}) can be roughly 
explained as follows.\cite{wenigerArXiv}
On supposing that our Dingle' series can be summed, the 
$m$th-order partials sum is written as the generalized limit (the ``sum'' 
of the series), say $T^{(k)}$, plus a remainder, 
say $R^{(k)}_m$, i.e.,  $S^{(k)}_m=T^{(k)}+R^{(k)}_m$, for any 
index $m$. For a diverging series, is just the divergent character of the sequence $\{R^{(k)}_m\}$ 
that ``hides'' the finite limit $T^{(k)}$. Thus the challenge 
consists in decoding the value of the limit from
the exploding behavior of $\{S^{(k)}_m\}$. The strategy common to 
most nonlinear sequence transformations, like WT,  consists in trying to 
compute an accurate approximation, say  $\bar R^{(k)}_m$, to the remainder $R^{(k)}_m$, by conceiving a suitable
asymptotic model, in order for $\bar R^{(k)}_m$ to be eliminated from 
$S^{(k)}_m$, which should provide  a better approximation 
$\{S^{(k)}_m-\bar R^{(k)}_m\}$ to the limit $T^{(k)}$  than $\{S^{(k)}_m\}$ itself.
For the Euler series, and thus for several factorial divergent series characterized by the asymptotic 
law in Eq.~(\ref{resurgence.2}), such an optimal remainder model has been found,\cite{wenigerCPR-89,wenigerArXiv} 
and leads to the WT in Eq.~(\ref{weniger}). 

In the final part of the present Letter we present some numerical 
tests devoted to illustrate the performances of WT in comparison to H 
and also to put into evidence some limitations of WT when it is  used 
in asymptotics.
In particular, we have chosen a class of phase integrals of pivotal importance 
in optics, namely the so-called \emph{cuspoid diffraction catastrophes},\cite{berryPIO-80}  for which the 
phase function is a $n$th-order polynomial of the form $\Phi(s)={s^{n}}+\sum_{k=1}^{n-2} x_k\,{s^{k}}$, with 
$\{x_1,x_2,...,x_{n-2}\}$ being a vector of (possibly complex) parameters.
Our example concerns with the asymptotics evaluation of the Pearcey function, corresponding to $n=4$. 
Furthermore, we also let $x_1=x$ and $x_2=y$. The hypeasymptotic evaluation of the Pearcey 
function was used, as a preliminary numerical test, in the 
original paper by BH.\cite{berryPRSA-91} Quite recently,  Davis and Kaminski\cite{parisJCAM-06}
have also proposed an hyperasymptotics computational scheme, based on 
Hadamard expansions, to retrieve very highly accurate evaluation of the 
Pearcey function for real values of $x$ and $y$.
There is also another reason for using such example to 
test the asymptotics of WT. Recently,\cite{borghiOL-07} the two simplest 
examples of diffraction catastrophes, the Airy ($n=3$) and just the Pearcey functions, have been 
computed via WT, but starting from their ascending power series. Thus the results we 
are going to present will be useful for comparison between the use of WT as far as 
Taylor-like expansions and asymptotics expansions are concerned. 
The evaluation of the coefficients in Eq.~(\ref{Dingle.2}) can be 
exactly in closed-form, and it is reported, for the Pearcey function, 
in Ref.~\onlinecite{berryPRSA-91} in terms of Gegenbauer 
polynomials.\cite{notaErrore}
\begin{figure}[!ht]
\centerline{\psfig{figure=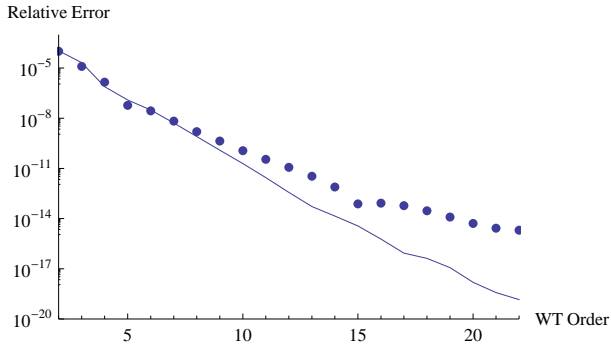,width=8cm,clip=,angle=0}}
\caption{Behavior of the relative error achieved by the WT, versus the order 
$m$ [see Eq.~(\ref{weniger})], when the Pearcey function is evaluated through the WT at the point
$(x=7,y=\sqrt 2(1+\mathrm{i}))$ (dots) and at the point $(x=-7,y=1)$ (solid curve).}
\label{fig1}
\end{figure}
Figure~\ref{fig1} shows the behavior of the relative error achieved by the WT, versus the order 
$m$ [see Eq.~(\ref{weniger})], when the Pearcey function is evaluated through  WT at the point
$(x=7,y=\sqrt 2(1+\mathrm{i}))$ (dots) and at the point $(x=-7,y=1)$ (solid curve).
The first case corresponds to that presented by BH in their original paper,\cite{berryPRSA-91} while the second one has been treated by Paris and Kaminski.\cite{parisJCAM-06} The two cases differ by 
the number of saddles which contribute to the Pearcey integral, respectively
one (complex) and three (reals). It is seen that in both cases the 
obtained accuracies are comparable to (and even better of) those obtained via the hyperasymptotics treatment.
In particular, in the first case WT provides a relative error of the order of $10^{-15}$ for $m=20$,
which is about 5 degree of magnitude better of that provided, on equal terms, by H.\cite{nota3}
The ultimate relative error obtained via H (see Tab.~1 in Ref.~\onlinecite{berryPRSA-91})
is of the order of $10^{-12}$, obtained with $m=25$, whereas WT provides, on equal terms, 
an error of $10^{-16}$ i.e., four degree of magnitude better.
Moreover, in Ref.~\onlinecite{borghiOL-07} the same case was treated 
by using WT  directly on the ascending  power series of the Pearcey 
function, and it turned out that to achieve a relative error of $10^{-10}$  
about  90 terms had to be considered in the partial sums sequence, whereas the same degree of accuracy is
here reached with less than 10 terms. The example treated by Paris and Kaminski corresponds to the physical situation of a
point located inside the cusp-shaped caustics, where the intensity 
distribution associated to the wavefield displays a typical fringe pattern, and
WT produces results which are as accurate as those find in 
Ref.~\onlinecite{parisJCAM-06} (see Tab.~1 of that paper). 

Before concluding the Letter, we want to give some warnings about 
possible limitations concerning the use of WT in asymptotics. We 
refer, in particular, to the fact that WT is not able, in the form 
given in Eq.~(\ref{weniger}), to resum nonalternating factorial divergent series,
like the Euler series in Eq.~(\ref{resurgence.2.1.1}) 
when $z$ is real and positive. \cite{nonalternating}
The above pathological situation occurs, for instance, when the Airy 
function Ai$(z)$  is required at the so-called Stokes lines, 
corresponding to complex value of $z$ such that 
$|\mathrm{arg}(z)|=2\pi/3$,\cite{berry234} where, on the contrary, the retrieving 
capabilities of H basically remain unchanged. We thus expect, in the  general 
case, that the effectiveness of WT will depend on the position of the complex vector $\{x_1,x_2,...,x_{n-2}\}$  with respect
to the  Stokes lines and surfaces.~\cite{wrightJPA-80,berry204} In this perspective,
a deeper investigation about the retrieving capabilities of WT is required, 
but the preliminary results here presented seem to corroborate our initial feeling
that Weniger transformation would be of great usefulness for  the asymptotic treatment of 
several optical problems.


\begin{thebibliography}{00}
\bibitem{notaErrore} The formula we are referring is Eq.~(42) of Ref.~\onlinecite{berryPRSA-91}, but the 
minus sign should be removed.
\bibitem{nota3} We are referring to
Fig.~12 of Ref.~\onlinecite{berryPRSA-91}.
\end{thebibliography}
\end{document}